\documentclass[conference,10pt]{IEEEtran}

\usepackage[dvips]{graphicx}
\usepackage{float}
\usepackage{times}
\usepackage{amsmath}
\usepackage{amsthm}
\usepackage{amsfonts}
\usepackage{gensymb}
\usepackage{cite}
\usepackage{bm}
\renewcommand{\baselinestretch}{1.05}

\begin{document}
%
\title{Study of Four-Dimensional DOA and Polarisation Estimation with Crossed-dipole and Tripole Arrays}


\author{\IEEEauthorblockN{Xiang Lan and Wei Liu }
	\IEEEauthorblockA{ 
		Department of Electronic and Electrical Engineering\\
		University of Sheffield\\
		Sheffield, S1 4ET, UK\\
{\it Email: \{xlan2, w.liu\}@sheffield.ac.uk}}
	\and
	\IEEEauthorblockN{Henry Y.T. Ngan}
	\IEEEauthorblockA{Department of Mathematics\\ Hong Kong Baptist University\\Kowloon Tong, Hong Kong\\
{\it Email: ytngan@hkbu.edu.hk}}
}

\maketitle

\begin{abstract}
Electromagnetic (EM) vector sensor arrays can track both the polarisation and direction of arrival (DOA) of the impinging signals. For linear crossed-dipole arrays, as shown by our analysis, due to inherent limitation of the structure, it can only track one DOA parameter and two polarisation parameters. For full four-dimensional (4-D, 2 DOA and 2 polarization parameters) estimation, we could extend the linear crossed-dipole array to the planar case. In this paper, instead of extending  the array geometry, we replace the crossed-dipoles by tripoles and construct a linear tripole array. It is proved that such a structure can estimate the 2-D DOA and 2-D polarisation information effectively in general and a dimension-reduction based MUSIC algorithm is developed so that the 4-D estimation problem can be simplified into two separate 2-D estimation problems, significantly reducing the computational complexity of the solution. The Cram\'{e}r-Rao Bound (CRB) is also derived as a reference for algorithm performance. A brief comparison between the planar crossed-dipole array and the linear tripole array is performed at last, showing that although the planar structure has a better performance, it is achieved at the cost of increased physical size.


\end{abstract}


%
\IEEEpeerreviewmaketitle

\section{Introduction}
The joint estimation of direction of arrival (DOA) and polarisation for signals based on electromagnetic (EM) vector sensor arrays has been widely studied in the past \cite{nehorai1994,shuai09,lan2017,yuan12,liu13n,liu14e,chevalier07,fried92,wong01,xu04,zoltowski00a,ferrara83,friedlander94,lee94,liu16c,liu16d,
li93,ye09,cadzow89,wax1985,he13,hurtado07,li1993,liu15a,liu16g}. In \cite{nehorai1994}, the EM vector sensor was first used to collect both electric and magnetic information of the impinging signals, where all six electromagnetic components are measured to identify the signals. So far most of the studies are focused on the linear array structure employing crossed-dipoles, where the general two-dimensional (2-D) DOA model is simplified into one-dimensional (1-D) by assuming that all the signals arrive from the same known azimuth angle $\phi$. In \cite{miron2006}, a quaternion MUSIC algorithm was proposed to deal with the joint DOA ($\theta$) and polarisation ($\rho$, $\phi$) estimation problem by considering the two complex-valued signals received by each crossed-dipole sensor as the four elements of a quaternion, where a three-dimensional (3-D) peak search is required with a very high computational complexity. In \cite{gong2008}, a quaternion ESPRIT algorithm was developed for direction finding with a reduced complexity. Furthermore, a dimension-reduction MUSIC algorithm based on uniform linear arrays (ULAs) with crossed-dipole sensors was introduced in \cite{zhang2014b}, where the 3-D joint peak search is replaced by a 1-D DOA search and a 2-D polarisation search.

In practice, the azimuth angle $\theta$ and the elevation angle $\phi$ of the signals are unknown and they are usually different for different signals and need to be estimated together. The existing 3-D joint DOA and polarisation work could be extended to 4-D (2 DOA and 2 polarisation). However, the 4-D estimation work comes with a uniqueness problem \cite{ho95,ho96,tan96,ho98,wong2000,tan96l}. In \cite{ho95}, it indicates that the uniqueness estimation problem is due to the linear dependence of joint steering vectors. In \cite{tan96l} and \cite{tan96}, Tan proved that for an EM vector sensor and EM vector sensor array, every three joint steering vectors with different DOAs are linearly independent, while the fourth joint steering vector with different DOA is possible to be the linear combination of the first three steering vectors. The linear dependence of steering vectors with tripole sensors is discussed in \cite{ho98} from the point of DOA estimation, where a special case of linear dependence is introduced that with some strict constraints, two steering vectors with different DOAs may be parallel to each other. Besides, the work also illustrates that the parallel can be avoided if the signals are nonlinearly polarised and arrives strictly from a hemispherical space.

When further reducing the tripole sensor array to a cross-dipole sensor array, as rigorously proved for the first time in this work, linear crossed-dipole array has the parallel ambiguity problem in general cases, where the azimuth angle and the elevation angle of the impinging signals can not be uniquely identified and there could also be false peaks in the resultant spatial spectrum. To tackle this ambiguity problem, one solution is to extend the linear geometry to a two-dimensional (2-D) rectangular planar array, such as the uniform rectangular array (URA), at significant space cost, and one good example for this solution is the work presented in \cite{hua1993}, where based on a URA, a pencil-MUSIC algorithm is proposed to solve the full 4-D DOA and polarisation estimation problem. However, it is not always feasible to use the rectangular array as a solution due to space limit. On the other hand, it is possible to add one dipole to the crossed-dipole structure to form a tripole sensor and tripole sensor array has been proposed in the past for DOA estimation~\cite{lundback04,liu15d}. Therefore, as another solution, motivated by simultaneously simplifying the array structure and reducing the computational complexity, instead of extending the linear crossed-dipole array to a higher spatial dimension, we replace the crossed-dipoles by tripoles and construct a linear tripole array in our earlier conference publication for joint 4-D DOA and polarisation estimation for the first time~\cite{lan17}. Moreover, for the first time, we give a clear proof about why a linear tripole array can be used for 4-D joint DOA and polarisation estimation, while avoiding the ambiguity problem except for some special cases.

At the algorithm level, two MUSIC-like algorithms for the 4-D estimation problem are proposed. The first is a direct search in the 4-D space to locate the DOA and polarisation parameters simultaneously (4-D MUSIC), which has an extremely high computational complexity. The other algorithm is to transform the 4-D search into two separate 2-D searches (2-D MUSIC), significantly reducing the computational complexity. To evaluate the performance of the proposed algorithms, the Cram\'{e}r-Rao Bound (CRB) of the linear tripole array for 4-D estimation is derived. In the past, CRBs have been derived under different circumstances, such as the results for arrays with arbitrary geometries in \cite{van2004}. Obviously, the types of signals and noise will affect the derived CRB result. Normally, noise is assumed to be temporally and spatially white and the source signal can have two different types: one is to assume the source signal is deterministic \cite{stoica1989,stoica1990m,bresler1986}, while the other assumes that the signal is random and a common choice is being Gaussion distributed \cite{abeida2017,kumar2015,lv2015,jin2009,roemer2006,stoica2001}. In this work, we assume the source signal is of the second type.

As mentioned, a URA of cross-dipoles can also achieve effective 4-D joint DOA and polarisation estimation. Then it would be interesting to know that given the same number of dipoles, which structure performs better. Our simulation results show that the planar array has a better performance, but this is achieved at the cost of increased physical size of the structure.

Overall, the contribution of our work is twofold. One is the detailed analysis and proof to show that the crossed-dipole linear array cannot uniquely identify the azimuth angle and the elevation angle of the impinging signals, while the tripole linear array can avoid the ambiguity problem for 4-D joint DOA and polarisation estimation in the general case. The other one is the proposed 4-D joint DOA and polarisation estimation method and its low-complexity version, with their performances compared to the newly derived CRB.

One note is that it is possible to have directive sensors with different orientations
to solve this ambiguity problem associated with the crossed-dipole array, but in general no fast search algorithms exist for such cases. Moreover,
the choice of orientation distribution will be another difficult problem to solve.

This paper is structured as follows. The linear tripole array is introduced in Section II with a detailed proof for the 4-D ambiguity problem associated with the linear crossed-dipole array and why the linear tripole array can solve the problem. The two 4-D estimation algorithms are proposed in Section III with the CRB derived in detail. Simulation results are presented in Section IV, and conclusions are drawn in Section V.

\section{Tripole Sensor Array Model}\label{sec:QM}
\subsection{Tripole sensor array}
\begin{figure}
  \centering
  \includegraphics[width=0.4\textwidth]{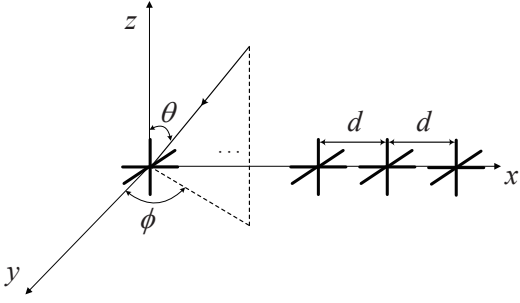}
  \caption{Geometry of a uniform linear tripole array, where a signal arrives from elevation angle $\theta$ and azimuth angle $\phi$.}\label{fig:01}
\end{figure}

Suppose there are $M$ uncorrelated narrowband signals impinging upon a uniform linear array with $N$ tripoles, where each tripole consists of three co-located mutually perpendicular dipoles, as shown in Fig. \ref{fig:01}. Assume that all signals are stationary and nonlinearly-polarised (elliptically or circularly polarised). The parameters, including DOA and polarisation of the $m$-th signal are denoted by $(\theta_{m},\phi_{m},\gamma_{m},\eta_{m}),m=1,2,...,M$, where $\theta_m\in[0,\pi/2]$, $\phi_m \in[0,2\pi]$, i.e. the signals come from the upper hemisphere. The inter-element spacing $d$ of the array is $\lambda/2$, where $\lambda$ is the wavelength of the incoming signals. For each tripole sensor, the three components are parallel to $x$, $y$ and $z$ axes, respectively. The background noise is Gaussian white with zero mean and variance $\sigma_n^{2}$, which is uncorrelated with the impinging signals. Due to the phase shift among the sensors, the steering vector for the $m$-th signal can be denoted as
\begin{eqnarray}
    \emph{\textbf{a}}_{m}=[1,e^{-j\pi \sin\theta_m\sin\phi_m},...,e^{-j(N-1)\pi \sin\theta_m\sin\phi_m}]
    \label{eq:steer1}
\end{eqnarray}
and the polarisation vector $\emph{\textbf{p}}_{m}$ is determined by the product of DOA component $\mathbf{\Omega}_{m}$ and the polarization component $\emph{\textbf{g}}_m$, where
\begin{eqnarray}
    \emph{\textbf{p}}_{m} = \mathbf{\Omega}_{m}\emph{\textbf{g}}_m
    \label{eq:polar}
\end{eqnarray}
The DOA component is a matrix consisting of two vectors that are orthogonal to the signal direction. There are infinite number of choices for these two vectors and generally the following two are used \cite{nehorai1998}
\begin{eqnarray}
        \mathbf{\Omega}_{m} =
                            \left[
                                \begin{matrix}
                                    \cos\theta_m\cos\phi_m & -\sin\phi_m \\
                                    \cos\theta_m\sin\phi_m & \cos\phi_m \\
                                    -\sin\theta_m & 0
                                \end{matrix}
                            \right]
\end{eqnarray}
The corresponding polarization component is given by
\begin{eqnarray}
    \emph{\textbf{g}}_{m} = \left[
                                \begin{matrix}
                                    \sin\gamma_m e^{j\eta_m} \\
                                    \cos\gamma_m
                                 \end{matrix}
                            \right]
\end{eqnarray}
where $\gamma_m$ is the auxiliary polarization angle and $\eta_m$ the polarization phase difference. By expanding (\ref{eq:polar}), the polarisation vector $\emph{\textbf{p}}_{m}$ can be divided into three different components in $x$, $y$ and $z$ axes
\begin{align}
    \emph{\textbf{p}}_{m} &=
                            \left[
                                \begin{matrix}
                                    \cos\theta_m\cos\phi_m\sin\gamma_m e^{j\eta_m}-\sin\phi_m\cos\gamma_m \\
                                    \cos\theta_m\sin\phi_m\sin\gamma_m e^{j\eta_m}+\cos\phi_m\cos\gamma_m \\
                                    -\sin\theta_m\sin\gamma_m e^{j\eta_m}
                                \end{matrix}
                            \right]
                            \label{eq:pm}
\end{align}
For convenience, we replace the three elements in $\emph{\textbf{p}}_{m}$ by $p_{mx}$, $p_{my}$ and $p_{mz}$, given by:
\begin{align}
    p_{mx}&=\cos\theta_m\cos\phi_m\sin\gamma_m e^{j\eta_m}-\sin\phi_m\cos\gamma_m \nonumber\\
    p_{my}&=\cos\theta_m\sin\phi_m\sin\gamma_m e^{j\eta_m}+\cos\phi_m\cos\gamma_m \nonumber\\
    p_{mz}&=-\sin\theta_m\sin\gamma_m e^{j\eta_m}
    \label{eq:xyz}
\end{align}
The received signal at the tripole sensor array can be denoted as a function of steering vector $\emph{\textbf{a}}_{m}$, polarisation vector $\emph{\textbf{p}}_{m}$, source signals $s_{m}(t)$ and background noise $\emph{\textbf{n}}$. At the $k$-th time instant, the received signal vector $\emph{\textbf{x}}[k]$ can be expressed as
\begin{eqnarray}
    \emph{\textbf{x}}[k] &=& \sum_{m=1}^{M}[\emph{\textbf{a}}_{m} \otimes \emph{\textbf{p}}_{m}]s_{m}[k]+\emph{\textbf{n}}[k]\nonumber\\
    &=& \sum_{m=1}^{M}\emph{\textbf{v}}_{m}s_{m}[k]+\emph{\textbf{n}}[k]
    \label{eq:steer}
\end{eqnarray}
where $\otimes$ stands for the Kronecker product, $\emph{\textbf{v}}_m$ is the Kronecker product of $\emph{\textbf{a}}_{m}$ and $\emph{\textbf{p}}_{m}$, and $\emph{\textbf{n}}[k]$ is the $3N\times1$ Gaussian white noise vector.
The covariance matrix $\emph{\textbf{R}}$ of the received signal vector is given by
\begin{eqnarray}
    \emph{\textbf{R}}&=&E\{\emph{\textbf{x}}[k]\emph{\textbf{x}}[k]^H\}\nonumber\\
    &=&\sum_{m=1}^{M}\emph{\textbf{v}}_{m}E\{s[k]s[k]^{*}\}\emph{\textbf{v}}_{m}^{H}+\sigma_n^2\emph{\textbf{I}}_{3N}
    \label{eq:cov}
\end{eqnarray}
In practice, $\emph{\textbf{R}}$ is not available and can be estimated by averaging a finite number of snapshots. In such a case, an estimated covariance matrix $\hat{\emph{\textbf{R}}}$ is used to replace $\emph{\textbf{R}}$
\begin{eqnarray}
    \hat{\emph{\textbf{R}}} \approx \frac{1}{K}\sum_{l=1}^{L}\emph{\textbf{x}}[k]\emph{\textbf{x}}[k]^H
\end{eqnarray}
where $K$ is the number of snapshots.
\subsection{Comparison between Crossed-dipole Array and Tripole Array}
This section will mainly show why the ULA with crossed-dipoles cannot uniquely determine the four parameters associated with each impinging signal, leading to the spatial aliasing problem, and why the ULA with tripoles can provide a unique solution for the joint 4-D estimation problem.

To show the ambiguity problem, we consider one source signal impinging upon the array so that the subscript $m$ can be dropped for convenience. The joint DOA and polarisation estimation problem can be considered as an estimation of the steering vector of this source signal.

For crossed-dipole sensor array, its joint steering vector $\emph{\textbf{w}}$ is given by
\begin{align}
    \emph{\textbf{w}} = \emph{\textbf{a}} \otimes \emph{\textbf{q}}
    \label{eq:cro}
\end{align}
where
\begin{align}
    \emph{\textbf{q}} =
                            \left[
                                \begin{matrix}
                                    p_x \\
                                    p_y
                                \end{matrix}
                            \right]
                            \label{eq:q}
\end{align}
Here, $\emph{\textbf{w}}$ is a $2N\times1$ vector with a $2\times1$ polarisation vector $\emph{\textbf{q}}$. For the tripole sensor array, the joint steering vector $\emph{\textbf{v}}$ is a $3N\times1$ vector with a $3\times1$ polarisation vector $\emph{\textbf{p}}$, i.e.
\begin{align}
    \emph{\textbf{v}} = \emph{\textbf{a}} \otimes \emph{\textbf{p}}
\end{align}
where
\begin{align}
    \emph{\textbf{p}}=    \left[
                                \begin{matrix}
                                    p_{x} \\
                                    p_{y} \\
                                    p_{z}
                                \end{matrix}
                            \right]
\end{align}

For convenience, we use $\alpha=(\theta,\phi,\gamma,\eta)$ to denote the four parameters. The ambiguity problem associated with the cross-dipole array can be stated as follows: If there is an arbitrarily polarised signal from $\alpha_1$, we can always find another signal from $\alpha_2$ that satisfies $\emph{\textbf{w}}_1//\emph{\textbf{w}}_2$, with $\alpha_1\neq\alpha_2$, where // means the two vectors are in parallel. By parallel we mean
\begin{align}
    \emph{\textbf{w}}_2 = k\cdot\emph{\textbf{w}}_1
\end{align}
where $k$ is an arbitrary complex-valued scalar.

When we say that the tripole array can avoid the ambiguity problem, it means that for nonlinearly polarised signals if $\alpha_1\neq\alpha_2$, the joint steering vectors $\emph{\textbf{v}}_1$ and $\emph{\textbf{v}}_1$ will never be in parallel with each other.

To prove these two statements, firstly we give the following definition and lemma.
\newtheorem*{mydef}{Definition}

\begin{mydef}
    Given two signals from distinct directions $(\theta_1,\phi_1)$ and $(\theta_2,\phi_2)$, the two signals are in DOA parallel if $\emph{\textbf{a}}_1=\emph{\textbf{a}}_2$.
\end{mydef}

Equation (\ref{eq:steer1}) indicates that $\emph{\textbf{a}}$ is only determined by the value of $\sin\theta\sin\phi$. If it satisfies that
\begin{align}
    \sin\theta_1\sin\phi_1=\sin\theta_2\sin\phi_2
    \label{eq:phi1}
\end{align}
the two steering vectors will be the same, i.e. $\emph{\textbf{a}}_1=\emph{\textbf{a}}_2$. It can be seen that in the upper hemisphere space ($0\leq \theta \leq \pi/2$, $0\leq \phi \leq 2\pi$), there are infinite number of directions in DOA parallel with a given direction.

\newtheorem*{lemma}{Lemma}
\begin{lemma}
    Given two complex-valued vectors $\emph{\textbf{w}}_1=\emph{\textbf{a}}_1 \otimes \emph{\textbf{q}}_1$ and $\emph{\textbf{w}}_2=\emph{\textbf{a}}_2 \otimes \emph{\textbf{q}}_2$, $\emph{\textbf{w}}_1//\emph{\textbf{w}}_2$ is necessary and sufficient for $\emph{\textbf{a}}_1//\emph{\textbf{a}}_2$ and $\emph{\textbf{q}}_1//\emph{\textbf{q}}_2$.
\end{lemma}

The proof of the lemma can be found in Appendix \ref{sectiona}.

Although we used the joint steering vector of the crossed-dipole array in the proof, it is straightforward to show that the lemma is also applicable to the joint steering vector of tripole sensor arrays.

Now we first consider the ambiguity problem in crossed-dipole sensor arrays. Given $\emph{\textbf{w}}_1=\emph{\textbf{a}}_1 \otimes \emph{\textbf{q}}_1$, our aim is to find a vector $\emph{\textbf{w}}_2=\emph{\textbf{a}}_2 \otimes \emph{\textbf{q}}_2$ with $\emph{\textbf{a}}_1//\emph{\textbf{a}}_2$ and $\emph{\textbf{q}}_1//\emph{\textbf{q}}_2$ when $\alpha_1\neq\alpha_2$.

As mentioned in the DOA parallel definition, any direction that satisfies (\ref{eq:phi1}) has the steering vector $\emph{\textbf{a}}_2//\emph{\textbf{a}}_1$. With the constraints, we need further choose values for $\gamma_2$ and $\eta_2$ to satisfy $\emph{\textbf{q}}_1//\emph{\textbf{q}}_2$. From (\ref{eq:xyz}) and (\ref{eq:q}), the polarisation vector $\emph{\textbf{q}}_1$ is determined by all four parameters $\theta_1$, $\phi_1$, $\gamma_1$ and $\eta_1$, where
\begin{align}
    \emph{\textbf{q}}_{1} &=
                            \left[
                                \begin{matrix}
                                    \cos\theta_1\cos\phi_1 & -\sin\phi_1 \\
                                    \cos\theta_1\sin\phi_1 & \cos\phi_1
                                \end{matrix}
                            \right]
                            \left[
                                \begin{matrix}
                                    \sin\gamma_1 e^{j\eta_1} \\
                                    \cos\gamma_1
                                 \end{matrix}
                            \right] \nonumber\\
                         &= \mathbf{\Psi}_{1}\emph{\textbf{g}}_1
\end{align}
Hence, the other polarisation vector $\emph{\textbf{q}}_2 = \mathbf{\Psi}_{2}\emph{\textbf{g}}_2$ needs to satisfy
\begin{align}
    &\mathbf{\Psi}_{1}\emph{\textbf{g}}_1 = \lambda \mathbf{\Psi}_{2}\emph{\textbf{g}}_2 \nonumber\\
    \Rightarrow &\emph{\textbf{g}}_2 = \lambda^{-1}\mathbf{\Psi}_{2}^{-1}\mathbf{\Psi}_{1}\emph{\textbf{g}}_1
\end{align}
$\lambda$ is a constant and without loss of generality we assume its value is 1. Here $\emph{\textbf{g}}_2$ is a $2\times1$ vector with $\emph{\textbf{g}}_2[1]=\sin\gamma_2 e^{j\eta_2}$ and $\emph{\textbf{g}}_2[2]=\cos\gamma_2$, where ``[1]" and ``[2]" denote the first and the second elements of the vector.
\begin{align}
    \tan\gamma_2 &= \frac{|\emph{\textbf{g}}_2[1]|}{|\emph{\textbf{g}}_2[2]|}\nonumber\\
    \tan\eta_2 &= \frac{\text{Im}\{\emph{\textbf{g}}_2[1]/\emph{\textbf{g}}_2[2]\}}
    {\text{Re}\{\emph{\textbf{g}}_2[1]/\emph{\textbf{g}}_2[2]\}}
    \label{eq:gamma1}
\end{align}
The new parameters from (\ref{eq:phi1}) ensure $\emph{\textbf{a}}_1//\emph{\textbf{a}}_2$ and the new parameters from (\ref{eq:gamma1}) ensure $\emph{\textbf{q}}_1//\emph{\textbf{q}}_2$ with the constraint $\alpha_1\neq\alpha_2$. After that, the new joint steering vector $\emph{\textbf{w}}_2$ will be in parallel with the original $\emph{\textbf{w}}_1$. As a result, we can not uniquely determine the four DOA and polarisation parameters of a source using the crossed-dipole array.

Next, we consider the tripole sensor array case. Given a joint steering vector $\emph{\textbf{v}}_1=\emph{\textbf{a}}_1 \otimes \emph{\textbf{p}}_1$, we want to prove that a parallel $\emph{\textbf{v}}_2=\emph{\textbf{a}}_2 \otimes \emph{\textbf{p}}_2$ does not exist and we prove it by contradiction. Similar to the crossed-dipole case, firstly a new direction which is in DOA parallel to the original direction is selected so that the new elevation and azimuth angles ensure $\emph{\textbf{a}}_1//\emph{\textbf{a}}_2$. This step is clearly feasible and the new direction can be obtained by (\ref{eq:phi1}). The remaining part of the problem is that whether there exists another polarisation vector $\emph{\textbf{p}}_2$ which is in parallel with $\emph{\textbf{p}}_1$. Assuming that $\emph{\textbf{p}}_2$ exists, i.e.
\begin{align}
    \mathbf{\Omega}_{1}\emph{\textbf{g}}_1 &= \lambda \mathbf{\Omega}_{2}\emph{\textbf{g}}_2
    \label{eq:omega}
\end{align}
where $\lambda$ is an unknown complex-valued constant. Expanding $\mathbf{\Omega}_{1}$ and $\mathbf{\Omega}_{2}$ by the column vector, where $\mathbf{\Omega}_{11}$ and $\mathbf{\Omega}_{12}$ are the first and second column vectors of $\mathbf{\Omega}_{1}$, and $\mathbf{\Omega}_{21}$ and $\mathbf{\Omega}_{22}$ are the first and second column vectors of $\mathbf{\Omega}_{2}$, respectively. (\ref{eq:omega}) is transformed to
\begin{align}
    [\mathbf{\Omega}_{11}\quad\mathbf{\Omega}_{12}]
                            \left[
                                \begin{matrix}
                                    \emph{\textbf{g}}_1[1] \\
                                    \emph{\textbf{g}}_1[2]
                                 \end{matrix}
                            \right]
                            &= \lambda [\mathbf{\Omega}_{21}\quad\mathbf{\Omega}_{22}]
                            \left[
                                \begin{matrix}
                                    \emph{\textbf{g}}_2[1] \\
                                    \emph{\textbf{g}}_2[2]
                                 \end{matrix}
                            \right]\nonumber\\
                            &\Updownarrow \nonumber\\
    \mathbf{\Omega}_{11}\emph{\textbf{g}}_1[1]+\mathbf{\Omega}_{12}\emph{\textbf{g}}_1[2]
    &=\mathbf{\Omega}_{21}\emph{\textbf{g}}_2[1]\lambda
    +\mathbf{\Omega}_{22}\emph{\textbf{g}}_2[2]\lambda
    \label{eq:lc}
\end{align}
The left side of (\ref{eq:lc}) can be viewed as a vector which is a linear combination of $\mathbf{\Omega}_{11}$ and $\mathbf{\Omega}_{12}$. The right is a linear combination of $\mathbf{\Omega}_{21}$ and $\mathbf{\Omega}_{22}$. Here we define a two-dimensional space $\emph{\textbf{A}}_1$ spanned by $\mathbf{\Omega}_{11}$ and $\mathbf{\Omega}_{12}$, also $\emph{\textbf{A}}_2$ spanned by $\mathbf{\Omega}_{21}$ and $\mathbf{\Omega}_{22}$. Since $\mathbf{\Omega}_{11},\mathbf{\Omega}_{12},\mathbf{\Omega}_{21}$ and $\mathbf{\Omega}_{22}$ are all $3\times1$ vectors, the equation holds only in the following two cases:

Case 1: $\emph{\textbf{A}}_1$ and $\emph{\textbf{A}}_2$ are the same two-dimensional span.

It can be noticed that $\emph{\textbf{A}}_1$ intersects with the $x-y$ plane at vector $\mathbf{\Omega}_{12}$, and $\emph{\textbf{A}}_2$ intersects with the $x-y$ plane at vector $\mathbf{\Omega}_{22}$. If $\emph{\textbf{A}}_1$ and $\emph{\textbf{A}}_2$ are the same two-dimensional span, it must satisfy that $\mathbf{\Omega}_{12}//\mathbf{\Omega}_{22}$, then we have
\begin{align}
\left[
    \begin{matrix}
    -\sin\phi_1 \\
    \cos\phi_1 \\
    0
    \end{matrix}
\right]//
\left[
    \begin{matrix}
    -\sin\phi_2 \\
    \cos\phi_2 \\
    0
    \end{matrix}
\right]
&\Leftrightarrow
-\frac{\sin\phi_1}{\cos\phi_1}=-\frac{\sin\phi_2}{\cos\phi_2}\nonumber\\
&\Leftrightarrow
\tan\phi_1 = \tan\phi_2
\label{eq:conflict}
\end{align}
However, $\phi_1\neq\phi_2$ and (\ref{eq:conflict}) conflicts with the basic assumption, which means that with the tripole sensor array, there is no other joint steering vector $\emph{\textbf{v}}_2$ in parallel with the given $\emph{\textbf{v}}_1$ in such a case.

Case 2: $\emph{\textbf{A}}_1$ and $\emph{\textbf{A}}_2$ are two different two-dimensional spans. Then $\emph{\textbf{p}}_1$ and $\emph{\textbf{p}}_2$ must be in parallel with the intersecting vector of $\emph{\textbf{A}}_1$ and $\emph{\textbf{A}}_2$.

Firstly we denote the intersecting vector as $\mathbf{\Omega}_x$. Since $\mathbf{\Omega}_{11},\mathbf{\Omega}_{12},\mathbf{\Omega}_{21},\mathbf{\Omega}_{22}$ are all real-valued vectors, all the elements in the intersection vector $\mathbf{\Omega}_x$ must also be real-valued. From eq.(\ref{eq:pm}), $\emph{\textbf{p}}_1$ can be transformed to
\begin{align}
    \emph{\textbf{p}}_{1} &=e^{j\eta}\left[
                                \begin{matrix}
                                    \cos\theta\cos\phi\sin\gamma -\sin\phi\cos\gamma e^{-j\eta} \\
                                    \cos\theta\sin\phi\sin\gamma +\cos\phi\cos\gamma e^{-j\eta} \\
                                    -\sin\theta\sin\gamma
                                \end{matrix}
                            \right]\nonumber\\
                          &=e ^{j\eta}\cdot\hat{\emph{\textbf{p}}}_{1}
\end{align}

It can be seen that $\emph{\textbf{p}}_{1}//\hat{\emph{\textbf{p}}}_{1}$. In most situations, with $\gamma\neq90\degree$, $\gamma\neq0$ and $\eta\neq0$ (nonlinearly polarized), the first two elements in $\hat{\emph{\textbf{p}}}_{1}$ are complex-valued and the last element in $\hat{\emph{\textbf{p}}}_{1}$ is real-valued, which indicates that with such a situation, it is impossible for $\hat{\emph{\textbf{p}}}_{1}$ to be in parallel with the intersecting vector $\mathbf{\Omega}_x$. Hence, if the incoming signal is nonlinearly polarised, for example, circular polarised or elliptically polarised, there is no ambiguity in joint estimation with tripole sensors. A detailed analysis about the ambiguity induced by linearly polarisation can be found in Appendix \ref{sectionb}.

\section{The proposed algorithm}\label{sec:AL}

In the following, the proposed low-complexity joint 4-D DOA and polarisation estimation algorithm for tripole sensor arrays is introduced based on a subspace approach.
\subsection{Joint 4-D Search}
Firstly, by applying eigenvalue decomposition (EVD), the covariance matrix $\emph{\textbf{R}}$ can be decomposed into
\begin{eqnarray}
    \emph{\textbf{R}} = \emph{\textbf{R}}_{s} + \emph{\textbf{R}}_{n}
    =\sum_{k=1}^{3N}\lambda_k\emph{\textbf{u}}_k\emph{\textbf{u}}_k^{H}
    \label{eq:2}
\end{eqnarray}
where $\emph{\textbf{u}}_k$ is the k-th eigenvector and $\lambda_k$ is the corresponding eigenvalue (in descending order).
Furthermore, we can rewrite (\ref{eq:2}) into
\begin{eqnarray}
    \emph{\textbf{R}} = \emph{\textbf{U}}_s\mathbf{\Lambda_s}\emph{\textbf{U}}_s^{H}+\emph{\textbf{U}}_n\mathbf\Lambda_n\emph{\textbf{U}}_n^{H}
    \label{11}
\end{eqnarray}
where $\emph{\textbf{U}}_s=[\emph{\textbf{u}}_1, \emph{\textbf{u}}_2, \cdots, \emph{\textbf{u}}_M]$ and $\emph{\textbf{U}}_n=[\emph{\textbf{u}}_{M+1}, \emph{\textbf{u}}_{M+2}, \cdots, \emph{\textbf{u}}_{3N}]$ are the eigenvectors of the signal subspace and noise subspace, respectively. $\mathbf{\Lambda_s}$ and $\mathbf\Lambda_n$ are diagonal matrices holding the corresponding eigenvalues $\lambda_k$. As the rank of the noise subspace cannot be less than 1, the DOF (degree of freedom) of the algorithm is $3N-1$, which means the maximum number of signals that can be estimated is $3N-1$.

Clearly, the joint steering vector $\emph{\textbf{v}}_m$ is orthogonal to the noise subspace $\emph{\textbf{U}}_n$, i.e.
\begin{eqnarray}
    \emph{\textbf{U}}_n^{H}\emph{\textbf{v}}_m=\textbf{0}
    \label{eq:ortho}
\end{eqnarray}
or
\begin{eqnarray}
    \emph{\textbf{v}}_m^{H}\emph{\textbf{U}}_n\emph{\textbf{U}}_n^{H}\emph{\textbf{v}}_m=0
    \label{eq:0}
\end{eqnarray}

As a result, to find the DOA and polarisation parameters $(\theta_m,\phi_m,\gamma_m,\eta_m)$ of the $m$-th signal, we construction the following function with normalization
\begin{eqnarray}
    F(\theta,\phi,\gamma,\eta)=\frac{1}{\emph{\textbf{v}}^{H}
    \emph{\textbf{U}}_n\emph{\textbf{U}}_n^{H}\emph{\textbf{v}}}
    \label{eq:estimator}
\end{eqnarray}

The peaks in (\ref{eq:estimator}) indicate the DOA and polarisation information $(\theta,\phi,\gamma,\eta)$ for impinging signals.

\subsection{Proposed Algorithm}
The above MUSIC-type algorithm is based on direct 4-D peak search with an extremely large computational complexity. In the following, we transform the 4-D search process into two 2-D searches, significantly reducing the complexity of the solution.

First, we separate $\emph{\textbf{v}}_m$ into two components: one with DOA information $(\theta,\phi)$ only, while the other only contains the polarisation information $(\gamma,\eta)$. In this way, (\ref{eq:ortho}) can be changed to
\begin{eqnarray}
    \textbf{0}&=&\emph{\textbf{U}}_n^{H}[\emph{\textbf{a}}_m \otimes (\mathbf{\Omega}_{m}\emph{\textbf{g}}_m)]\nonumber\\
    &=&\emph{\textbf{U}}_n^{H}[(\emph{\textbf{a}}_m \otimes \mathbf{\Omega}_{m}) \emph{\textbf{g}}_m]\nonumber\\
    &=&[\emph{\textbf{U}}_n^{H}\emph{\textbf{B}}_m]\emph{\textbf{g}}_m
    \label{eq:gm}
\end{eqnarray}
where $\emph{\textbf{B}}_m$ is the Kronecker product of $\emph{\textbf{a}}_m$ and $\mathbf{\Omega}_{m}$.

Following the approach in \cite{ferrara83} for three-dimensional estimation (one DOA parameter and two polarisation parameters), an estimator can be constructed by searching for the minimum eigenvalue of the $2 \times 2$ matrix as follows,
\begin{eqnarray}
    f_1(\theta,\phi)=\frac{1}{\lambda_{min}\{\emph{\textbf{B}}^{H}\emph{\textbf{U}}_n\emph{\textbf{U}}_n^{H}\emph{\textbf{B}} \}}
    \label{eq:estimator2.1}
\end{eqnarray}
where $\lambda_{min}$ denotes the minimum eigenvalue of the matrix.

Note that $\emph{\textbf{U}}_n^{H}\emph{\textbf{B}}_m$ is a $(3N-M)\times2$ vector and $\emph{\textbf{g}}_m$ is a $2 \times 1$ vector. (\ref{eq:gm}) indicates that $\emph{\textbf{g}}_m$ lies in the null space of $\emph{\textbf{U}}_n^{H}\emph{\textbf{B}}_m$. Since $\emph{\textbf{U}}_n^{H}\emph{\textbf{B}}_m$ is a $(3N-M)\times2$ matrix, it has the null space only if its rank is less than or equivalent to 1. Consequently, multiplied by the Hermitian transpose on the right, the new $2 \times 2$ product matrix cannot have a full rank, which means the determinant equals to zero. Here we use $det\{\}$ to denote the determinant of a matrix. Then, we have
\begin{eqnarray}
    det\{\emph{\textbf{B}}_m^{H}\emph{\textbf{U}}_n\emph{\textbf{U}}_n^{H}\emph{\textbf{B}}_m\}=0
\end{eqnarray}

We can see that $\emph{\textbf{B}}_m$ is dependent on the parameters $(\theta,\phi)$ only. As a result, a new estimator can be established corresponding to $\theta$ and $\phi$ as \cite{chen2016}
\begin{eqnarray}
    f_2(\theta,\phi)=\frac{1}{det\{\emph{\textbf{B}}^{H}\emph{\textbf{U}}_n\emph{\textbf{U}}_n^{H}\emph{\textbf{B}} \}}
    \label{eq:estimator2}
\end{eqnarray}

Compared to the solution in (29) based on calculating the minimum eigenvalue, the determinant-based solution in (31) has a lower complexity as will be shown next in Sec. III-B, although there is no clear difference between their estimation performances  as demonstrated by computer simulations later. After performing a 2-D peak search over $\theta$ and $\phi$ by (\ref{eq:estimator2.1}) or (\ref{eq:estimator2}), the polarisation parameters $\gamma$ and $\eta$ can be obtained by another 2-D search in the following
\begin{eqnarray}
    f_2(\gamma,\eta)=\frac{1}{\emph{\textbf{g}}^{H}\emph{\textbf{B}}^{H}\emph{\textbf{U}}_n\emph{\textbf{U}}_n^{H}
    \emph{\textbf{B}}\emph{\textbf{g}}}
    \label{eq:estimator3}
\end{eqnarray}

The following is a summary of the proposed algorithm:
\begin{itemize}
  \item Calculate the estimated covariance matrix $\hat{\emph{\textbf{R}}}$ from the received signals.
  \item Calculate the noise space $\emph{\textbf{U}}_n$ by applying the eigenvalue decomposition on $\hat{\emph{\textbf{R}}}$. The last $3N-M$ eigenvalues and the corresponding eigenvectors form the noise space.
  \item Use the 2-D estimator (\ref{eq:estimator2}) to locate the DOA parameters $\theta$ and $\phi$.
  \item Use the 2-D estimator (\ref{eq:estimator3}) to locate the corresponding polarisation parameters $\gamma$ and $\eta$.
\end{itemize}

\subsection{Complexity Comparison}

For all the algorithms (the solution in (\ref{eq:estimator}), the solution in (\ref{eq:estimator2.1}) and the solution in (\ref{eq:estimator2}) together with their associated solution in (\ref{eq:estimator3})), they have the same process of calculating the covariance matrix $\emph{\textbf{R}}$ and its EVD. Therefore, to compare their computational complexity, we ignore this common part and focus on the complexity of the searching process. In the following analysis, the number of search points of each parameter is assumed to be the same, which is $L$.

In the direct 4-D search algorithm, the four parameters are estimated within $L^4$ searches. During each search, it requires $3N(3N-M)$ multiplications and $(3N-1)(3N-M)$ additions to calculate $\emph{\textbf{v}}^{H}\emph{\textbf{U}}_n$. The denominator in (\ref{eq:estimator}) is the product of $\emph{\textbf{v}}^{H}\emph{\textbf{U}}_n$ and its Hermitian transpose, which requires $(3N-M)$ multiplications and $(3N-M-1)$ additions. Consequently, if the additions are ignored, the 4-D search requires $L^4(3N+1)(3N-M)$ multiplications. In the 2-D search analysis, the additions will be ignored as well.

In the 2-D search algorithm, we discuss the computational complexity of eigenvalue based estimator (\ref{eq:estimator2.1}) and (\ref{eq:estimator2}) respectively. The two estimators can be both divided into the DOA estimation step and the polarisation estimation step. The complexity difference is in their DOA estimation step and the two estimators have the same complexity in the polarisation estimation step. In the following, the DOA estimation step will be firstly discussed.

The DOA parameters estimated by (\ref{eq:estimator2.1}) are within $L^2$ searches. During each search, the computation operations will be doubled to calculate $\emph{\textbf{B}}^{H}\emph{\textbf{U}}_n$ compared to the operations of $\emph{\textbf{v}}^{H}\emph{\textbf{U}}_n$ since $\emph{\textbf{B}}^{H}$ is a $2\times(3N-M)$ matrix. The required multiplications will be $6N(3N-M)$. The product of $\emph{\textbf{B}}^{H}\emph{\textbf{U}}_n$ and its Hermitian transpose requires four times operations as those needed for 4-D search, which is $4(3N-M)$ multiplications.
Besides, the computation of the minimum eigenvalue requires six multiplications. As a result, the step for DOA estimation by (\ref{eq:estimator2.1}) requires $L^2[(6N+4)(3N-M)+6]$ multiplications.

When using (\ref{eq:estimator2}) to estimate DOAs, during each search, it has the same complexity to calculate the matrix $\emph{\textbf{B}}^{H}\emph{\textbf{U}}_n\emph{\textbf{U}}_n^{H}\emph{\textbf{B}}$ with $(6N+4)(3N-M)$ multiplications. While the computation of the determinant of the $2\times2$ matrix requires only two multiplications. Thus, the complexity of DOA estimation by (\ref{eq:estimator2}) is $L^2[(6N+4)(3N-M)+2]$.

In the polarisation estimation step, the polarisation parameters are estimated by (\ref{eq:estimator3}) within another $L^2$ searches. As $\emph{\textbf{B}}^{H}$ has already been estimated by the first step, there is no need to calculate $\emph{\textbf{B}}^{H}\emph{\textbf{U}}_n$ in every search. Similar to the DOA search step, the product of $\emph{\textbf{B}}^{H}\emph{\textbf{U}}_n$ and its Hermitian transpose requires $(6N+4)(3N-M)$ multiplications. After that, the denominator matrix in (\ref{eq:estimator3}) requires eight multiplications. Hence, the multiplications required for the polarisation parameters will be $8L^2+(6N+4)(3N-M)$. To sum up, the 2-D search estimator by (\ref{eq:estimator2.1}) has a complexity of $L^2[(6N+4)(3N-M)+14]+(6N+4)(3N-M)$ and the estimator (\ref{eq:estimator2}) has a complexity of $L^2[(6N+4)(3N-M)+10]+(6N+4)(3N-M)$.

In practice, especially in high resolution estimations, $L$ is far larger than $M$ and $N$. The complexity of the two algorithms is mainly dependent on the value of $L$. Based on the given results, it can be seen that the 2-D search has a much lower computational complexity $O(L^2)$ than the direct 4-D search $O(L^4)$. Moreover, the determinant based 2-D search estimator has a slightly lower computational complexity than the eigenvalue based 2-D search estimator.

\subsection{Cram\'{e}r-Rao Bound for Tripole Sensor Array}
The Cram\'{e}r-Rao bound (CRB) provides a lower bound on the variance of unbiased estimators. In the joint estimation problem, ($\theta,\phi,\gamma,\eta$) are four unknown parameters. With the N-element linear tripole array, the probability density function for a single received snapshot is given by \cite{van2004}
\begin{eqnarray}
    p_{x}|(\alpha) = \frac{1}{\text{det}[\pi \emph{\textbf{R}}_{\emph{\textbf{x}}}(\alpha)]}e^{\{-[\emph{\textbf{x}}(t)-\emph{\textbf{m}}(\alpha)]^{H}
    \emph{\textbf{R}}_{\emph{\textbf{x}}}^{-1}(\alpha)[\emph{\textbf{x}}(t)-\emph{\textbf{m}}(\alpha)]\}}
\end{eqnarray}
where $\emph{\textbf{R}}_{\emph{\textbf{x}}}(\alpha)$ is the covariance matrix and $\emph{\textbf{m}}(\alpha)$ is the mean value of received vector data.

With $K$ independent snapshots, the likelihood function can be denoted as the product of $K$ single functions
\begin{align}
    p_{x_{1},x_{2},...,x_{K}}|(\alpha)=&\prod_{k=1}^{K}\frac{1}{\text{det}[\pi \emph{\textbf{R}}_{\emph{\textbf{x}}}(\alpha)]}\nonumber\\
    &\times e^{\{-[\emph{\textbf{x}}_{k}-\emph{\textbf{m}}(\alpha)]^{H}
    \emph{\textbf{R}}_{\emph{\textbf{x}}}^{-1}(\alpha)[\emph{\textbf{x}}_{k}-\emph{\textbf{m}}(\alpha)]\}}
\end{align}

The log-likelihood function is given by
\begin{align}
    L_{x}(\alpha)=&\ln{p_{x_{1},x_{2},...,x_{K}}|(\alpha)}\nonumber\\
    =&-K\ln{\text{det}[\emph{\textbf{R}}_{\emph{\textbf{x}}}(\alpha)]}-KN\ln{\pi}\nonumber\\
    &-\sum_{k=1}^{K}{[\emph{\textbf{x}}_{k}-\emph{\textbf{m}}(\alpha)]^{H}
    \emph{\textbf{R}}_{\emph{\textbf{x}}}^{-1}(\alpha)[\emph{\textbf{x}}_{k}-\emph{\textbf{m}}(\alpha)]}
    \label{eq:likelihood}
\end{align}

Considering the unconditional model, which means the source signals are random in all realizations \cite{stocia1990}, for one source the mean value and covariance matrix are given by
\begin{align}
    \emph{\textbf{m}}(\alpha)&=\emph{\textbf{0}}\nonumber\\
    \emph{\textbf{R}}_{\emph{\textbf{x}}}(\alpha)&=E[\emph{\textbf{x}}(t)-\emph{\textbf{m}}(\alpha)]^{H}[\emph{\textbf{x}}(t)-\emph{\textbf{m}}(\alpha)]\nonumber\\
    &= E[\emph{\textbf{x}}^{H}(t)\emph{\textbf{x}}(t)]
    = \sigma_s^2{\emph{\textbf{v}}\emph{\textbf{v}}^{H}}+\sigma_n^2\emph{\textbf{I}}
    \label{eq:umodel}
\end{align}
where $\sigma_s^2$ is the power of source signal and $\sigma_n^2$ is the noise power.

The Fisher information matrix can be denoted as:
\begin{align}
    \emph{\textbf{F}}(\alpha)=\left[\begin{matrix}
                                        F_{\theta,\theta} \quad F_{\theta,\phi} \quad F_{\theta,\gamma} \quad F_{\theta,\eta}\\
                                        F_{\phi,\theta} \quad F_{\phi,\phi} \quad F_{\phi,\gamma} \quad F_{\phi,\eta}\\
                                        F_{\gamma,\theta} \quad F_{\gamma,\phi} \quad F_{\gamma,\gamma} \quad F_{\gamma,\eta}\\
                                        F_{\eta,\theta} \quad F_{\eta,\phi} \quad F_{\eta,\gamma} \quad F_{\eta,\eta}
                                    \end{matrix}
                                \right]
\end{align}
Each element in the matrix can be expressed as the product of derivatives of (\ref{eq:likelihood}) with respect to the corresponding parameter \cite{van2004}:
\begin{align}
    F_{\alpha_{i},\alpha_{j}}=&\text{tr}\{\emph{\textbf{R}}_{\emph{\textbf{x}}}^{-1}(\alpha)\frac{\partial \emph{\textbf{R}}_{\emph{\textbf{x}}}(\alpha)}{\alpha_{i}}\emph{\textbf{R}}_{\emph{\textbf{x}}}^{-1}(\alpha)\frac{\partial \emph{\textbf{R}}_{\emph{\textbf{x}}}(\alpha)}{\alpha_{j}}\}\nonumber\\
    &+2Re\{\frac{\partial\emph{\textbf{m}}^{H}(\alpha)}{\alpha_{i}}\emph{\textbf{R}}_{\emph{\textbf{x}}}^{-1}(\alpha)\frac{\partial\emph{\textbf{m}}(\alpha)}{\alpha_{j}}\}
    \label{eq:fishe}
\end{align}
where the symbol $\text{tr}\{\}$ denotes the trace of a matrix, $Re\{\}$ the real part, and $\alpha_{i},\alpha_{j}$ two arbitrary parameters among ($\theta,\phi,\gamma,\eta$).

With (\ref{eq:umodel}), (\ref{eq:fishe}) can be simplified to
\begin{align}
    F_{\alpha_{i},\alpha_{j}}=\text{tr}\{\emph{\textbf{R}}_{\emph{\textbf{x}}}^{-1}(\alpha)\frac{\partial \emph{\textbf{R}}_{\emph{\textbf{x}}}(\alpha)}{\alpha_{i}}\emph{\textbf{R}}_{\emph{\textbf{x}}}^{-1}(\alpha)\frac{\partial \emph{\textbf{R}}_{\emph{\textbf{x}}}(\alpha)}{\alpha_{j}}\}
    \label{eq:fishe2}
\end{align}
The CRB matrix $\emph{\textbf{C}}(\alpha)$ is the inverse of Fisher information matrix, i.e.
\begin{align}
    \emph{\textbf{C}}(\alpha)=\emph{\textbf{F}}^{-1}(\alpha)
\end{align}
Finally, the Cram\'{e}r-Rao bounds for each estimated parameter are given by:
\begin{align}
    CRB(\theta)&=C_{\theta,\theta}=[\emph{\textbf{F}}^{-1}(\alpha)]_{1,1}\nonumber\\
    CRB(\phi)&=C_{\phi,\phi}=[\emph{\textbf{F}}^{-1}(\alpha)]_{2,2}\nonumber\\
    CRB(\gamma)&=C_{\gamma,\gamma}=[\emph{\textbf{F}}^{-1}(\alpha)]_{3,3}\nonumber\\
    CRB(\eta)&=C_{\eta,\eta}=[\emph{\textbf{F}}^{-1}(\alpha)]_{4,4}
\end{align}
\section{Simulation Results}
In this section, simulation results are presented to demonstrate the ambiguity issues discussed earlier and the performance of the proposed algorithm.
\subsection{Ambiguity Phenomenon}
Assuming one source signal from $(\theta,\phi,\gamma,\eta)=(30\degree,80\degree,20\degree,50\degree)$ impinges on a uniform linear crossed-dipole array and a uniform linear tripole array respectively. Both arrays have the same senor number $N=5$ and the inter-element space is set to $d=\lambda/2$.

\begin{figure}
  \centering
  \includegraphics[width=0.4\textwidth]{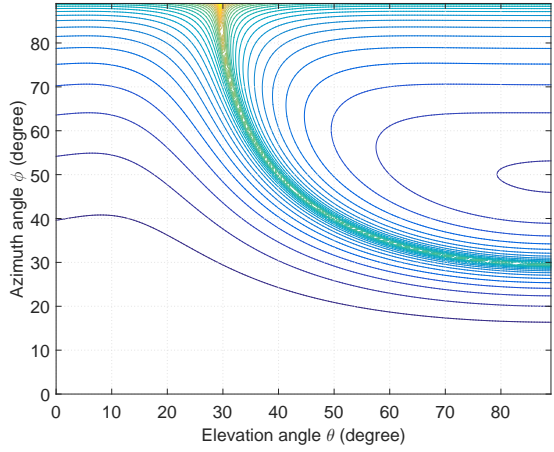}
  \caption{DOA estimation spectrum using the linear crossed-dipole sensor array (top contour view).}\label{fig:04}
  \vspace{1ex}
\end{figure}

\begin{figure}
  \centering
  \includegraphics[width=0.4\textwidth]{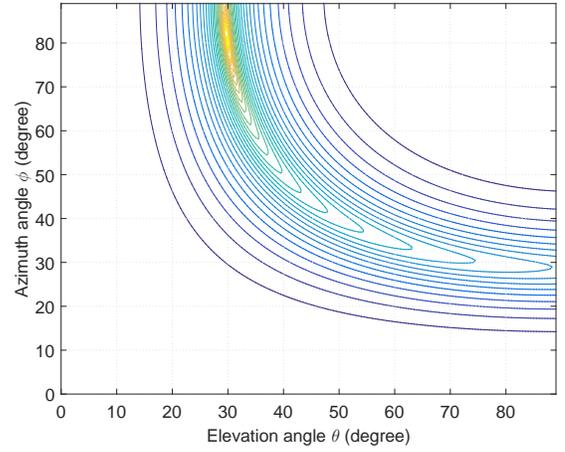}
  \caption{DOA estimation spectrum using the linear tripole sensor array (top contour view).}\label{fig:05}
  \vspace{1.5ex}
\end{figure}

Figs. \ref{fig:04} and \ref{fig:05} present the DOA estimation results for these two arrays, respectively. Apparently, the tripole array gives a unique peak at the source direction while the crossed-dipole array shows a peak line due to the ambiguity problem and there is no way to identify the real direction of the signal.
\subsection{RMSE Results}

\begin{figure}
  \centering
  \includegraphics[width=0.4\textwidth]{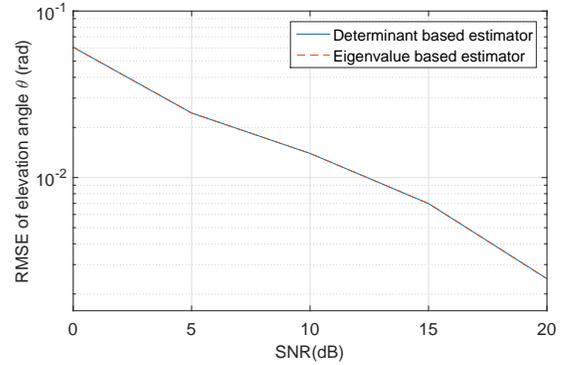}
  \caption{RMSE of $\theta$ with determinant based and eigenvalue based estimators.}\label{fig:t1}
\end{figure}

\begin{figure}
  \centering
  \includegraphics[width=0.4\textwidth]{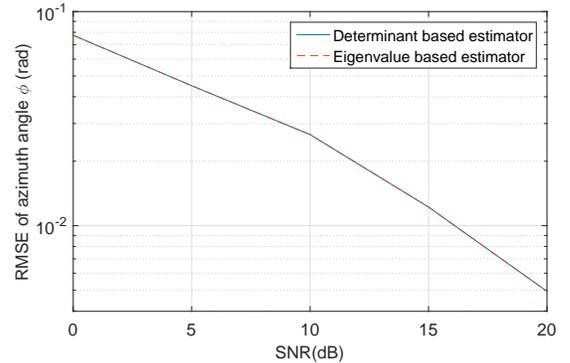}
  \caption{RMSE of $\phi$ with determinant based and eigenvalue based estimators.}\label{fig:t2}
\end{figure}

Now we study the performance of the proposed algorithm based on tripole sensor arrays. Firstly, we make a comparison between the performance of the two 2-D estimators in (\ref{eq:estimator2.1}) and (\ref{eq:estimator2}). Consider a single source signal from $(\theta,\phi,\gamma,\eta)=(10\degree,20\degree,15\degree,30\degree)$ impinging on a ULA with four tripole sensors and half-wavelength spacing. The root mean square error (RMSE) versus SNR of the two 2-D estimators are plotted with snapshot number $K=1000$ and 200 Monte-Carlo trials in Figs.\ref{fig:t1} and \ref{fig:t2}. From the results, it can be observed that the two estimators have no clear difference in their performance.

\begin{figure}
  \centering
  \includegraphics[width=0.4\textwidth]{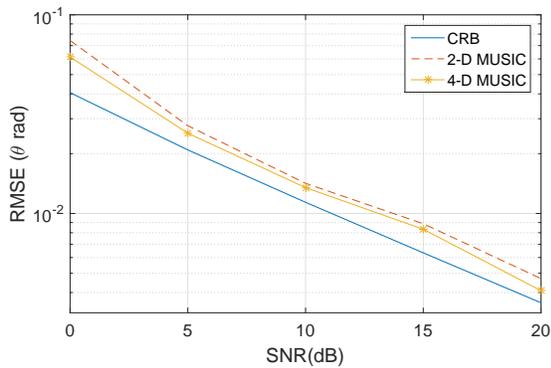}
  \caption{RMSE of $\theta$.}\label{fig:06}
  \vspace{2ex}
\end{figure}

\begin{figure}
  \centering
  \includegraphics[width=0.4\textwidth]{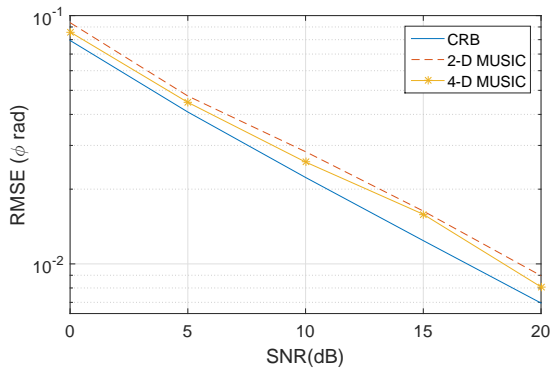}
  \caption{RMSE of $\phi$.}\label{fig:07}
  \vspace{2ex}
\end{figure}

\begin{figure}
  \centering
  \includegraphics[width=0.4\textwidth]{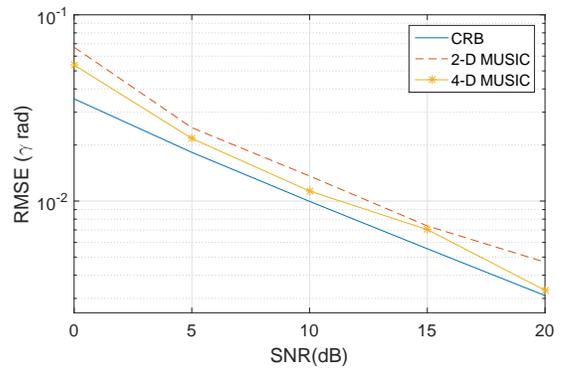}
  \caption{RMSE of $\gamma$.}\label{fig:08}
  \vspace{2ex}
\end{figure}

\begin{figure}
  \centering
  \includegraphics[width=0.4\textwidth]{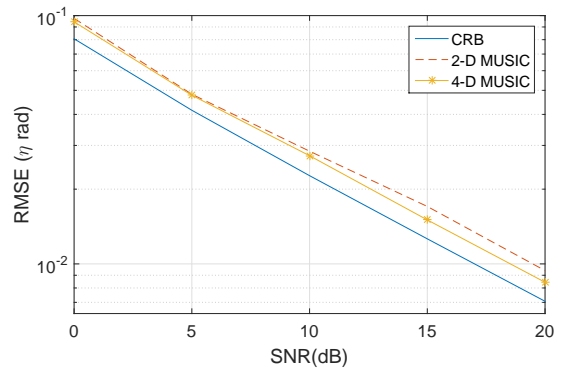}
  \caption{RMSE of $\eta$.}\label{fig:09}
  \vspace{2ex}
\end{figure}

As the two 2-D estimators have the same performance, in the following, we only compare the determinant based 2-D estimator with the 4-D estimator and CRB. Assume there are two source signals from $(\theta,\phi,\gamma,\eta)=(10\degree,20\degree,15\degree,30\degree)$ and $(\theta,\phi,\gamma,\eta)=(60\degree,70\degree,60\degree,80\degree)$. The tripole sensor number is set to $N=4$ and the number of snapshots for each simulation is $K=1000$. The RMSE results of the estimated parameters by 200 Monte-Carlo trials are shown in Figs. \ref{fig:06}-\ref{fig:09}, where we can see that with the increase of SNR, the RMSE level decreases consistently. The accuracy of the 4-D MUSIC using (\ref{eq:estimator}) is always better than the proposed 2-D MUSIC algorithm for any parameters at the cost of a much higher level of computational complexity. The performance of both algorithms are close to the CRB.
\subsection{Linear Tripole and Planar Crossed-dipole Array}
Since a planar crossed-dipole array can also be used to estimate the four parameters of an impinging signal, it would be interesting to know that given the same number of dipoles, which one is more effective for 4-D parameter estimation, the linear tripole array or the planar crossed-dipole array. To find out, in this part, we consider a $4\times1$ linear tripole array and a $2\times3$ planar crossed-dipole array both of which have the same number of dipoles or DOFs. We compare their estimation accuracy using the proposed 2-D MUSIC algorithm. All the other conditions are the same as in Section IV-B.
\vspace{1ex}

\begin{figure}
  \centering
  \includegraphics[width=0.4\textwidth]{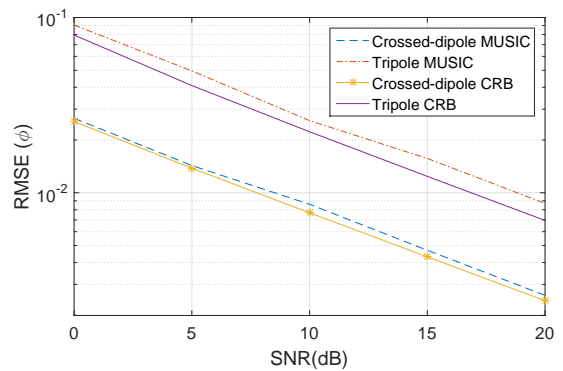}
  \caption{RMSE of crossed-dipole and tripole sensor array.}\label{fig:10}
\end{figure}
Fig. \ref{fig:10} shows the RMSE results for the first signal's azimuth angle. It can be seen that the planar array has given a higher estimating accuracy and its CRB is much lower than the linear tripole array, which means that the compact structure of the linear tripole sensor array is achieved at the cost of estimation accuracy.
\section{Conclusion and Discussion}

With a detailed analysis and proof, it has been shown that due to inherent limitation of the linear crossed-dipole structure, it cannot uniquely identify the four parameters associated with impinging signals. In order to simultaneously estimate both the 2-D DOA and 2-D  polarisation parameters of the impinging signals, we could increase the dimension of the array and construct a planar crossed-dipole array. To avoid this and have a compact structure, a linear tripole array has been employed instead. It has been proved and also shown that such a structure can estimate the 2-D DOA and 2-D polarisation information effectively except for some very special cases. Moreover, a dimension-reduction based MUSIC algorithm was developed so that the 4-D estimation problem can be simplified to two separate 2-D estimation problems, significantly reducing the computational complexity of the solution. However, the dimension-reduction also brings the problem of less accuracy. Since both the planar crossed-dipole array and the linear tripole array can be used to effectively estimate the four parameters of an impinging signal, a brief comparison between them was also carried out and it was shown that given the same number of dipoles, the planar structure has a better performance, although this is achieved at the cost of increased physical size.

\appendices
\section{Proof of the Lemma}\label{sectiona}

Necessity: If $\emph{\textbf{a}}_1//\emph{\textbf{a}}_2$ and $\emph{\textbf{q}}_1//\emph{\textbf{q}}_2$, then
\begin{align}
    \emph{\textbf{a}}_2 = k_1\cdot\emph{\textbf{a}}_1 \nonumber\\
    \emph{\textbf{q}}_2 = k_2\cdot\emph{\textbf{q}}_1
\end{align}
where $k_1$ and $k_2$ are arbitrary complex-valued constants.
Then,
\begin{align}
    \emph{\textbf{w}}_2 &= \emph{\textbf{a}}_2 \otimes \emph{\textbf{q}}_2\nonumber\\
                        &= (k_1\cdot\emph{\textbf{a}}_1) \otimes (k_2\cdot\emph{\textbf{q}}_1) \nonumber\\
                        &= (k_1k_2)\cdot(\emph{\textbf{a}}_1 \otimes \emph{\textbf{q}}_1) \nonumber\\
                        &= (k_1k_2)\cdot\emph{\textbf{w}}_1
\end{align}
Hence, $\emph{\textbf{w}}_1//\emph{\textbf{w}}_2$.

Sufficiency: By (\ref{eq:cro}), $\emph{\textbf{w}}$ can be expanded as
\begin{align}
    \emph{\textbf{w}} = \emph{\textbf{a}} \otimes \emph{\textbf{q}}
                      = \left[
                                \begin{matrix}
                                    a_{1}\emph{\textbf{q}} \\
                                    .\\
                                    .\\
                                    .\\
                                    a_{N}\emph{\textbf{q}}
                                \end{matrix}
                            \right]
                          = \left[
                                \begin{matrix}
                                    a_{1}p_x \\
                                    a_{1}p_y \\
                                    .\\
                                    .\\
                                    .\\
                                    a_{N}p_x\\
                                    a_{N}p_y
                                \end{matrix}
                            \right]
\end{align}
The Hermitian transpose $\emph{\textbf{w}}^H$ is given by
\begin{align}
    \emph{\textbf{w}}^H = \emph{\textbf{a}}^H \otimes \emph{\textbf{q}}^H
                        \label{eq:Herv}
\end{align}
The norm of $\emph{\textbf{w}}$ is
\begin{align}
    |\emph{\textbf{w}}| &= \sqrt{\emph{\textbf{w}}^H\emph{\textbf{w}}}\nonumber\\
                        &= \sqrt{(a_1a_1^*+...+a_Na_N^*)(p_x p_x^*+p_y p_y^*)}\nonumber\\
                        &=|\emph{\textbf{a}}|\cdot|\emph{\textbf{q}}|
\end{align}
Then, we have
\begin{align}
    |\emph{\textbf{w}}_1| &= |\emph{\textbf{a}}_1|\cdot|\emph{\textbf{q}}_1|\nonumber\\
    |\emph{\textbf{w}}_2| &= |\emph{\textbf{a}}_2|\cdot|\emph{\textbf{q}}_2|
\end{align}
Generally, by (\ref{eq:Herv}), the modulus of the inner product of $\emph{\textbf{w}}_1$ and $\emph{\textbf{w}}_2$ can be expanded as
\begin{align}
    |\emph{\textbf{w}}_1^H\emph{\textbf{w}}_2| &= |(\emph{\textbf{a}}_1^H \otimes \emph{\textbf{q}}_1^H)\cdot(\emph{\textbf{a}}_2 \otimes \emph{\textbf{q}}_2)|
    \label{eq:mixpro}
\end{align}
According to the mixed-product property of Kronecker product, lemma 4.2.10 in \cite{roger94}, (\ref{eq:mixpro}) can be deduced to
\begin{align}
    |\emph{\textbf{w}}_1^H\emph{\textbf{w}}_2| &= |\emph{\textbf{a}}_1^H\cdot\emph{\textbf{a}}_2|\otimes|\emph{\textbf{q}}_1^H\cdot\emph{\textbf{q}}_2|\nonumber\\
    &\leq |\emph{\textbf{a}}_1|\cdot|\emph{\textbf{a}}_2|\cdot|\emph{\textbf{q}}_1|\cdot|\emph{\textbf{q}}_2|
    \label{eq:pro2}
\end{align}
On the other hand, since $\emph{\textbf{w}}_1//\emph{\textbf{w}}_2$, we know $\emph{\textbf{w}}_2=k\emph{\textbf{w}}_1$ and $|\emph{\textbf{w}}_2|=|k||\emph{\textbf{w}}_1|$, which leads to
\begin{align}
    |\emph{\textbf{w}}_1^H\emph{\textbf{w}}_2| &= |\emph{\textbf{w}}_1^H\cdot k\emph{\textbf{w}}_1|=|k||\emph{\textbf{w}}_1|\cdot|\emph{\textbf{w}}_1|\nonumber\\
    &=|\emph{\textbf{w}}_1|\cdot|\emph{\textbf{w}}_2|=|\emph{\textbf{a}}_1|\cdot|\emph{\textbf{a}}_2|\cdot|\emph{\textbf{q}}_1|\cdot|\emph{\textbf{q}}_2|
    \label{eq:pro1}
\end{align}

The equality in (\ref{eq:pro2}) holds only when $\emph{\textbf{a}}_1//\emph{\textbf{a}}_2$ and $\emph{\textbf{q}}_1//\emph{\textbf{q}}_2$. Combined with (\ref{eq:pro1}), the sufficiency proof is completed.

\section{Ambiguity on Tripole Sensor Array with Linearly Polarised Signals}\label{sectionb}
If the signals are linearly polarised, $\gamma=90\degree$ or $\gamma=0$ or $\eta=0$. $\hat{\emph{\textbf{p}}}_{1}$ or $\emph{\textbf{p}}_{1}$ becomes a vector with all elements being real-valued, and it may be possible for  $\emph{\textbf{p}}_{1}$ to be in parallel with the intersecting vector $\mathbf{\Omega}_x$. Now with the assumption $\emph{\textbf{p}}_{1}//\emph{\textbf{p}}_{2}//\mathbf{\Omega}_x$, $\emph{\textbf{p}}_{1}$ and $\emph{\textbf{p}}_{2}$ must all be real-valued, which means $\gamma_1=90\degree$ or $\gamma_1=0$ or $\eta_1=0$, and at the same time $\gamma_2=90\degree$ or $\gamma_2=0$ or $\eta_2=0$. With the constraint $\sin\theta_1\sin\phi_1=\sin\theta_2\sin\phi_2$, we consider all of the nine different cases:

Case 1: $\gamma_1=90\degree$ and $\gamma_2=90\degree$.

In this case
\begin{align}
    \emph{\textbf{p}}_{1} &=e^{j\eta_1}\left[
                                \begin{matrix}
                                    \cos\theta_1\cos\phi_1 \\
                                    \cos\theta_1\sin\phi_1 \\
                                    -\sin\theta_1
                                \end{matrix}
                            \right]\nonumber\\
    \emph{\textbf{p}}_{2} &=e^{j\eta_2}\left[
                                \begin{matrix}
                                    \cos\theta_2\cos\phi_2 \\
                                    \cos\theta_2\sin\phi_2 \\
                                    -\sin\theta_2
                                \end{matrix}
                            \right]
\end{align}
With $\theta_1=\theta_2$ and $\phi_1=\phi_2$, we have $\emph{\textbf{p}}_{1}//\emph{\textbf{p}}_{2}$ for arbitrary $\eta_1$ and $\eta_2$. An example is $(30\degree,60\degree,90\degree,20\degree)$ and $(30\degree,60\degree,90\degree,50\degree)$.

Case 2: $\gamma_1=90\degree$ and $\gamma_2=0\degree$. (same for $\gamma_1=0^\circ$ and $\gamma_2=90^\circ$)
\begin{align}
    \emph{\textbf{p}}_{1} &=e^{j\eta_1}\left[
                                \begin{matrix}
                                    \cos\theta_1\cos\phi_1 \\
                                    \cos\theta_1\sin\phi_1 \\
                                    -\sin\theta_1
                                \end{matrix}
                            \right]\nonumber\\
    \emph{\textbf{p}}_{2} &=\left[
                                \begin{matrix}
                                    -\sin\phi_2 \\
                                    \cos\phi_2 \\
                                    0
                                \end{matrix}
                            \right]
\end{align}
In this case, with $\theta_1=0\degree$ and $\tan\phi_1=-\cot\phi_2$, we have $\emph{\textbf{p}}_{1}//\emph{\textbf{p}}_{2}$ for arbitrary $\theta_2$, $\eta_1$ and $\eta_2$. An example is $(0\degree,90\degree,90\degree,20\degree)$ and $(50\degree,0\degree,0\degree,50\degree)$.

Case 3: $\gamma_1=90\degree$ and $\eta_2=0\degree$. (same for $\eta_1=0^\circ$ and $\gamma_2=90^\circ$)
\begin{align}
    \emph{\textbf{p}}_{1} &=e^{j\eta_1}\left[
                                \begin{matrix}
                                    \cos\theta_1\cos\phi_1 \\
                                    \cos\theta_1\sin\phi_1 \\
                                    -\sin\theta_1
                                \end{matrix}
                            \right]\nonumber\\
    \emph{\textbf{p}}_{2} &=\left[
                                \begin{matrix}
                                   \cos\theta_2\cos\phi_2\sin\gamma_2-\sin\phi_2\cos\gamma_2 \\
                                    \cos\theta_2\sin\phi_2\sin\gamma_2+\cos\phi_2\cos\gamma_2 \\
                                    -\sin\theta_2\sin\gamma_2
                                \end{matrix}
                            \right]
\end{align}
Given arbitrary $\theta_1,\phi_1,\theta_2,\phi_2$ which satisfy the constraint (\ref{eq:phi1}), if $\emph{\textbf{p}}_{1}//\emph{\textbf{p}}_{2}$, then
\begin{align}
    \left\{
\begin{aligned}
    \frac{\sin\theta_1}{\sin\theta_2\sin\gamma_2}=\frac{\cos\theta_1\cos\phi_1}{\cos\theta_2
    \cos\phi_2\sin\gamma_2-\sin\phi_2\cos\gamma_2}\\
    \frac{\sin\theta_1}{\sin\theta_2\sin\gamma_2}=\frac{\cos\theta_1\sin\phi_1}{\cos\theta_2
    \sin\phi_2\sin\gamma_2+\cos\phi_2\cos\gamma_2}
\end{aligned}
    \right.
\end{align}
leading to
\begin{align}
    \left\{
    \begin{aligned}
        \sin\phi_2&=\cos\phi_2\\
        \sin\phi_2&=-\cos\phi_2
    \end{aligned}
    \right.
\end{align}
which causes contradiction. In this case, there is no ambiguity.

Case 4: $\gamma_1=0\degree$ and $\gamma_2=0\degree$.
\begin{align}
    \emph{\textbf{p}}_{1} &=\left[
                                \begin{matrix}
                                    -\sin\phi_1 \\
                                    \cos\phi_1 \\
                                    0
                                \end{matrix}
                            \right]\nonumber\\
    \emph{\textbf{p}}_{2} &=\left[
                                \begin{matrix}
                                    -\sin\phi_2 \\
                                    \cos\phi_2 \\
                                    0
                                \end{matrix}
                            \right]
\end{align}
In this case, with $\phi_1=\phi_2$, we have $\emph{\textbf{p}}_{1}//\emph{\textbf{p}}_{2}$ for arbitrary $\eta_1$ and $\eta_2$. An example is $(30\degree,60\degree,0\degree,20\degree)$ and $(30\degree,60\degree,0\degree,50\degree)$.

Case 5: $\gamma_1=0\degree$ and $\eta_2=0\degree$. (same for $\eta_1=0^\circ$ and $\gamma_2=0^\circ$)
\begin{align}
    \emph{\textbf{p}}_{1} &=\left[
                                \begin{matrix}
                                    -\sin\phi_1 \\
                                    \cos\phi_1 \\
                                    0
                                \end{matrix}
                            \right]\nonumber\\
    \emph{\textbf{p}}_{2} &=\left[
                                \begin{matrix}
                                   \cos\theta_2\cos\phi_2\sin\gamma_2-\sin\phi_2\cos\gamma_2 \\
                                    \cos\theta_2\sin\phi_2\sin\gamma_2+\cos\phi_2\cos\gamma_2 \\
                                    -\sin\theta_2\sin\gamma_2
                                \end{matrix}
                            \right]
\end{align}
In this case, to satisfy the parallel condition, firstly $\theta_2$ should be $0\degree$ and $\eta_1$ can be an arbitrary value. Further we have
\begin{align}
    \tan\gamma_2=\frac{\cos\phi_1\sin\phi_2-\sin\phi_1\cos\phi_2}{\cos\phi_1\cos\phi_2+\sin\phi_1\sin\phi_2}
\end{align}
An example is $(30\degree,0\degree,0\degree,30\degree)$ and $(0\degree,30\degree,30\degree,0\degree)$.

Case 6: $\eta_1=0\degree$ and $\eta_2=0\degree$.
\begin{align}
    \emph{\textbf{p}}_{1} &=\left[
                                \begin{matrix}
                                    \cos\theta_1\cos\phi_1\sin\gamma_1-\sin\phi_1\cos\gamma_1 \\
                                    \cos\theta_1\sin\phi_1\sin\gamma_1+\cos\phi_1\cos\gamma_1 \\
                                    -\sin\theta_1\sin\gamma_1
                                \end{matrix}
                            \right]\nonumber\\
    \emph{\textbf{p}}_{2} &=\left[
                                \begin{matrix}
                                   \cos\theta_2\cos\phi_2\sin\gamma_2-\sin\phi_2\cos\gamma_2 \\
                                    \cos\theta_2\sin\phi_2\sin\gamma_2+\cos\phi_2\cos\gamma_2 \\
                                    -\sin\theta_2\sin\gamma_2
                                \end{matrix}
                            \right]
\end{align}
In this case, due to the parallel condition, we know
\begin{align}
    \left\{
\begin{aligned}
    \frac{\sin\theta_1\sin\gamma_1}{\sin\theta_2\sin\gamma_2}=\frac{\cos\theta_1\cos\phi_1\sin\gamma_1-\sin\phi_1
    \cos\gamma_1}{\cos\theta_2\cos\phi_2\sin\gamma_2-\sin\phi_2\cos\gamma_2}\\
    \frac{\sin\theta_1\sin\gamma_1}{\sin\theta_2\sin\gamma_2}=\frac{\cos\theta_1\sin\phi_1\sin\gamma_1+\cos\phi_1
    \cos\gamma_1}{\cos\theta_2\sin\phi_2\sin\gamma_2+\cos\phi_2\cos\gamma_2}
\end{aligned}
    \right.
    \label{eq:case6}
\end{align}
Each equations in (\ref{eq:case6}) will produce a unique solution to $\tan\gamma_2$. Except that all the parameters $(\theta_1,\phi_1,\gamma_1)=(\theta_2,\phi_2,\gamma_2)$, there is no other solutions for $\gamma_2$ and therefore there is no ambiguity in this case.

\bibliographystyle{IEEEtran}
\bibliography{mybib}

\end{document}